\documentclass[aps,prl,showpacs,floatfix,twocolumn,amsmath,amssymb,preprintnumbers]{revtex4-1}
\usepackage{mathrsfs}
\usepackage[figuresright]{rotating}
\usepackage{amsmath}
\usepackage{amssymb}
\usepackage{graphicx}
\usepackage{color}
\usepackage{dcolumn}
\usepackage{bm}
\usepackage[breaklinks=true,colorlinks=true,linkcolor=blue,urlcolor=blue,citecolor=blue]{hyperref}

\usepackage{soul}

\makeatletter

\newcommand{\Rmnum}[1]{\expandafter\@slowromancap\romannumeral #1@}
\makeatother

\begin{document}

\title{PrBi: Topology meets quadrupolar degrees of freedom}

\author{Xiaobo He$^{1}$}
\author{Chuanwen Zhao$^{2}$}
\author{Haiyang Yang$^{3}$}
\author{Jinhua Wang$^{1}$}
\author{Kangqiao Cheng$^{1}$}
\author{Shan Jiang$^{1}$}
\author{Lingxiao Zhao$^{1}$}
\author{Yuke Li$^{3}$}
\author{Chao Cao$^{3}$}
\author{Zengwei Zhu$^{1}$}
\author{Shun Wang$^{2}$}
\author{Yongkang Luo$^{1}$}
\email[]{mpzslyk@gmail.com}
\author{Liang Li$^{1}$}
\address{$^1$Wuhan National High Magnetic Field Center and School of Physics, Huazhong University of Science and Technology, Wuhan 430074, China;}
\address{$^2$MOE Key Laboratory of Fundamental Physical Quantities Measurement $\&$ Hubei Key Laboratory of Gravitation and Quantum Physics, School of Physics, Huazhong University of Science and Technology, Wuhan 430074, China;}
\address{$^3$Department of Physics and Hangzhou Key Laboratory of Quantum Matters, Hangzhou Normal University, Hangzhou 311121, China.}

\date{\today}

\begin{abstract}

Novel materials incorporating electronic degrees of freedom other than charge, including spin, orbital or valley \textit{et al} have manifested themselves to be of the great interests and applicable potentials. Recently, the multipolar degrees of freedom have attracted remarkable attention in the electronic correlated effects. In this work, we systematically studied the transport, magnetic and thermodynamic properties of the topological semimetal candidate PrBi in the framework of crystalline electric field theory. Our results demonstrate the $\Gamma_3$ non-Kramers doublet as the ground state of Pr$^{3+}$ (4$f^2$) ions. This ground state is nonmagnetic but carries a non-zero quadrupolar moment $\langle\hat{O}_2^0\rangle$. A quadrupolar phase transition is inferred below 0.08 K. No obvious quadrupolar Kondo effect can be identified. Ultrahigh-field quantum oscillation measurements confirm PrBi as a semimetal with non-trivial Berry phase and low total carrier density 0.06 /f.u. We discuss the interplay between low carrier density and $4f^2$ quadrupolar moment, and ascribe the weak quadrupolar ordering and Kondo effect to consequences of the low carrier density. PrBi, thus, opens a new window to the physics of topology and strongly correlated effect with quadrupolar degrees of freedom in the low-carrier-density limit, evoking the need for a reexamination of the Nozi\`{e}res exhaustion problem in the context of multi-channel Kondo effect.

\end{abstract}


\maketitle

\section{\Rmnum{1}. Introduction}

Kondo effects involving quadrupolar degrees of freedom have attracted extensive attention. A conventional Kondo effect\cite{Kondo-RMinimum,Hewson-Kondo} depicts that the magnetic moment (spin, \textit{i.e.} dipolar degrees of freedom) of a spin-1/2 impurity immersed into the sea of conduction ($c$) electrons is screened and quenched by forming an entangled Kondo singlet\cite{Coleman-HFDimension}. On the other hand, magnetic exchanges between these local moments can be realized via the so-called Ruderman-Kittel-Kasuya-Yosida (RKKY) interaction which is also mediated by the conduction electrons\cite{RKKY-RK,RKKY-K,RKKY-Y}. The competition between Kondo effect and RKKY interaction leads to a quantum critical point\cite{Doniach} in the vicinity of which many emergent quantum phenomena may appear, such as heavy-fermion, non-Fermi liquid, unconventional superconductivity and so on. Theories have also suggested that nonmagnetic Kondo effect can also be realized by taking the multipolar degrees of freedom. Cox proposed that $f^2$ ions like Pr$^{3+}$ or U$^{4+}$ sitting in a cubic-symmetrized crystalline electric field (CEF) may have $\Gamma_3$ nonmagnetic doublet ground state whose \textit{quadrupolar moment} can interact with conduction electrons\cite{Cox-Quadrupolar,Cox-PhysicaB1994}, termed as \textit{quadrupolar Kondo effect}. Likewise, indirect RKKY-like interaction was also proposed to play an essential role to mediate the long-range quadrupolar ordering\cite{Levy-QInderaction,Onimaru-PrPb3Modulated}. A variety of interesting properties associated with the quadrupolar degrees of freedom have been observed, see for examples the Pr-based intermetallic family Pr$Tm_2X_{20}$, where $Tm$=Ti,V,Rh,Ir, and $X$=Al,Zn\cite{Sato-PrTi2Al20INS,Onimaru-PrIr2Zn20AFQSC,Onimaru-PrRh2Zn20AFQSC,Sakai-PrTrAl20JPSJ,Tsujimoto-PrV2Al20SC,Matsubayashi-PrTi2Al20Pressure}.

For whatever dipolar or quadrupolar Kondo effect, a general presumption is that the system contains indefinitely sufficient conduction electrons to screen the local moments. An interesting question, then, is what if in a system with low carrier density? This encourages us to recall the famous Nozi\`{e}res exhaustion idea\cite{Nozieres-EPJB1998} that an insufficient number of conduction-electron spins to separately screen a large amount of local moments. In the dipolar Kondo compounds Yb$X$Cu$_4$ ($X$=Ag,Au,Cd,Mg,Tl,and Zn), this idea has been testified by the fact that the development of Kondo coherence is severely protracted as carrier density decreases\cite{Sarrao-YbXCu4,Lawrence-YbXCu4}; whereas the similar question remains open in the quadrupolar Kondo effect. Intuitively, carrier density should be even more critical for the quadrupolar Kondo effect, because here the Kondo effect is multi-channel and thus requires an \textit{over-screening} number of conduction electrons. The condition on the RKKY side is equally interesting. Because low carrier density means a dearth of conduction electrons to transfer the RKKY interaction, the long-range ordering is expected to be weakened, too. A further question one may pose is, on the whole, how does the low carrier density affect those emergent phenomena involved by quadrupolar degrees of freedom? To answer these questions, proper material basis with $f^2$ ions in cubic site symmetry, and most importantly, of low carrier density, are needed.

In this work, we systematically studied the physical properties of the cubic-structured semimetal PrBi. Previous studies on PrBi have reported a large magnetoresistance (MR) which potentially originates from electron-hole compensation\cite{VashistA-PrBiSdH,WuZ-PrSmSbBiPRB2019}. Herein, the quantum oscillation measurements up to 53 T not only confirm the low carrier density of $\sim$0.06 /f.u., but also point to a topologically non-trivial Berry phase. The magnetic and thermodynamic properties of PrBi are well understood in the framework of CEF theory. We demonstrate that the CEF ground state of Pr$^{3+}$ $4f^2$ electron is a nonmagnetic $\Gamma_3$ doublet whose quadrupolar moments likely form a quadrupolar ordering below 0.08 K. The influence of low carrier density on the correlation effect involving quadrupolar degrees of freedom is discussed.

\section{\Rmnum{2}. Experimental datails}

High quality single crystalline PrBi and its non-$4f$ reference LaBi were grown by Sb-flux method as described elsewhere\cite{VashistA-PrBiSdH,WuZ-PrSmSbBiPRB2019,NiuXH-LaBiARPES}. Transport measurements were made in a standard Hall-bar configuration in a commercial Physical Property Measurement System with a rotator option (PPMS-9 T, Quantum Design). Magnetic susceptibility is measured in a Magnetic Property Measurement System (MPMS, Quantum Design) equipped with a vibrating sample magnetometer (VSM). The specific heat measurements were performed by a thermal relaxation method down to 0.08 K in a Dilution Refrigerator insert of PPMS. Magnetoresistance was also measured under pulsed magnetic field up to 53 T at Wuhan National High Magnetic Field Center.

First-principles calculations based on density functional theory (DFT) were performed using the plane-wave basis projected augmented wave (PAW) method as implemented in Vienna Abinit Simulation Package (VASP)\cite{Kresse-VASP}. The Pr-$4f$ orbitals were regarded as local core-states that provide the quadrupole instead of valence electrons, and do not explicitly enter the calculations. The energy-cutoff was chosen to be 480 eV, and a 12$\times$12$\times$12 $\mathbf{\Gamma}$-centered $\mathbf{k}$-mesh was used to converge the total energy to 1 meV per unit cell. The modified Becke-Johnson (mBJ)\cite{Tran-mBJ} was employed to obtain the electronic structure, which is then fitted using the maximally localized Wannier function method\cite{Arash-mlWf} to obtain the Fermi surface properties as well as the quantum oscillation frequencies.

\section{\Rmnum{3}. Results and Discussion}

\begin{figure*}[t]
\vspace*{-5pt}
\hspace*{-10pt}
\includegraphics[width=18.5cm]{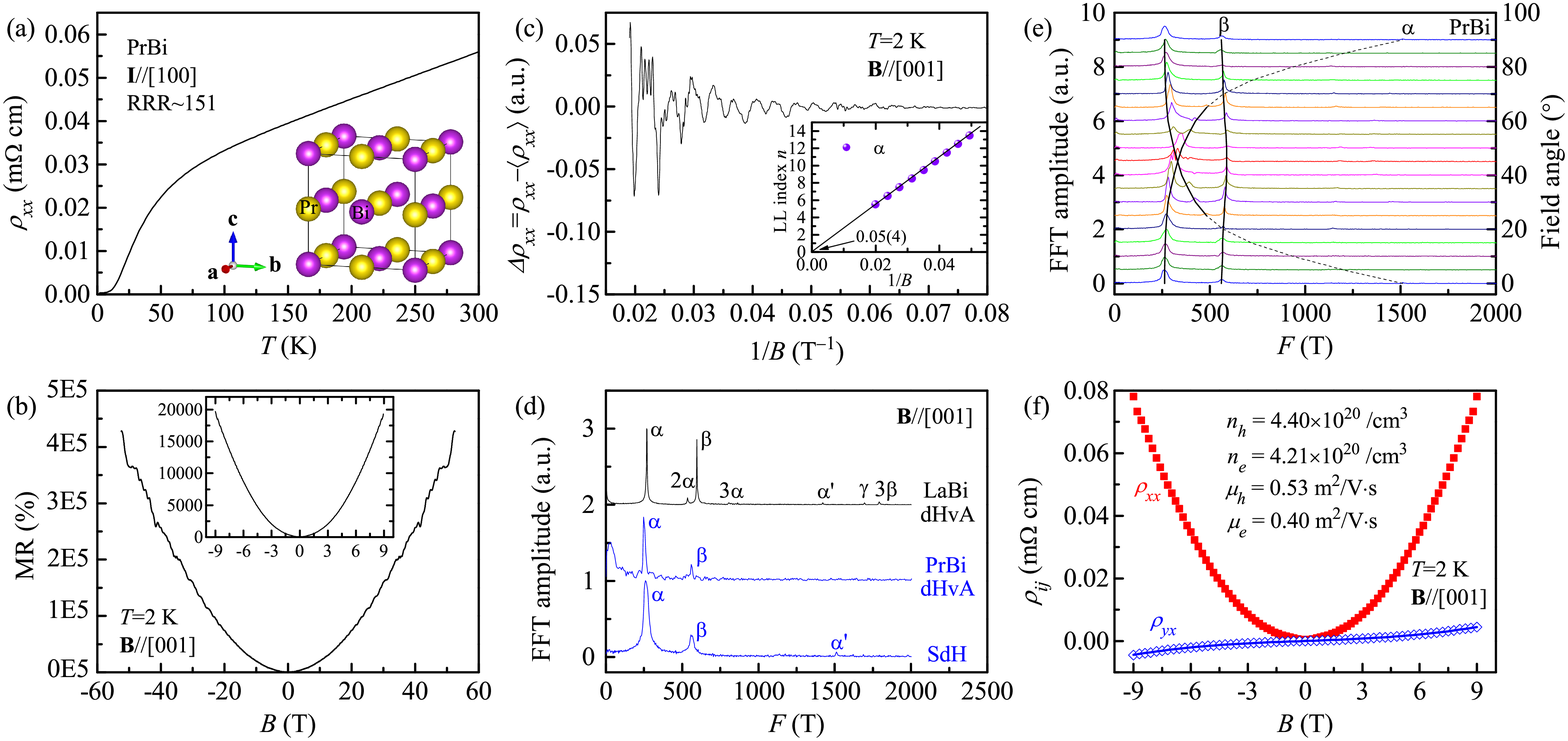}
\vspace*{-0pt}
\caption{\label{Fig1} Electrical transport properties of PrBi. (a) Resistivity as a function of temperature, $\rho_{xx}(T)$, in the absence of magnetic field. (b) MR reaches 430,000\% (20,000\%) for field up to 53 T (9 T). (c) SdH oscillation in $\Delta\rho_{xx}$ as a function of $1/B$. The inset depicts the Landau-level fan plot. The intercept at infinite-field limit indicates non-trivial Berry phase. (d) FFT of dHvA and SdH oscillations of PrBi (blue) and LaBi (black). (e) SdH FFT of PrBi for various field orientations, $\mathbf{B}$$\perp$$\mathbf{I}$. The black lines are guide lines to the Fermi surface structure. The dotted line is speculated from DFT calculations\cite{Hasegawa-LaBiFS}. (f) $\rho_{xx}(B)$ and $\rho_{yx}(B)$ measured at 2 K. The solid line represents the two-band fitting of $\rho_{yx}(B)$. }
\end{figure*}

The compound studied here, PrBi, is a topological semimetal with low carrier density. This can be seen from the transport measurements as shown in Fig.~\ref{Fig1}. The sample measured is of good quality, guaranteed by the large residual resistance ratio (RRR) about 151, comparable to previous reports\cite{VashistA-PrBiSdH,WuZ-PrSmSbBiPRB2019}. Over the full range 2-300 K, we did not see any anomaly or in particular any trace of $-\log T$ behavior in $\rho_{xx}(T)$ as seen in some Pr-based intermetallic compounds\cite{Sakai-PrTrAl20JPSJ}. This implies that quadrupolar Kondo effect arising from $c$-$f$ hybridization is weak for temperature above 2 K. Under magnetic field and at low temperature, it exhibits extremely large magnetoresistance, as commonly seen in many other semimetals\cite{Ali-WTe2XMR}. At 2 K, the MR reaches 440,000\% (20,000\%) for field up to 53 T (9 T). The nearly parabolic $\rho_{xx}(B)$ profile and unsaturated MR up to 53 T should be attributed to electron-hole compensation\cite{Ali-WTe2XMR,LuoY-WTe2Hall,VashistA-PrBiSdH,WuZ-PrSmSbBiPRB2019,DaiYM-WTe2UFOS}.

\begin{figure}[!htp]
\vspace*{-0pt}
\hspace*{-20pt}
\includegraphics[width=12cm]{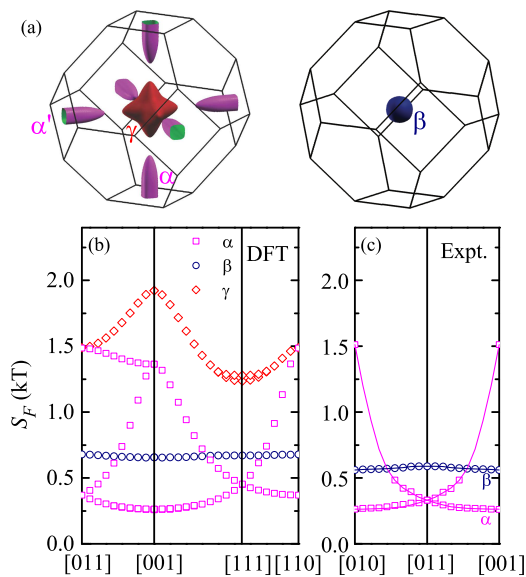}
\vspace*{-15pt}
\caption{\label{Fig2} Fermi surface of PrBi. (a) DFT calculated Fermi surface of PrBi. The Fermi surfaces are indexed as $\alpha$ (electron-type, ellipsoidal, at X), $\beta$ (hole-type, spherical, at $\Gamma$), and $\gamma$ (hole-type, star-shaped, at $\Gamma$), respectively. The angular dependence of Fermi surface cross-sections are shown in panels (b) and (c), from theoretical and experimental, respectively. }
\end{figure}

Clear Shubnikov-de Hass (SdH) quantum oscillations can be seen in $\rho_{xx}(B)$. After subtracting the non-oscillatory background $\langle\rho_{xx}\rangle$, we show the oscillatory part $\Delta\rho_{xx}$ in Fig.~\ref{Fig1}(c) as a function of $1/B$. The Fast Fourier Transform (FFT) pattern displays multiple frequencies, cf Fig.~\ref{Fig1}(d) where the de Haas-van Alphen (dHvA) oscillation results of LaBi and PrBi are also presented for comparison. The quantum oscillation arises from quantized Landau levels passing over the Fermi surface, and the frequency ($F$) of oscillation in the $1/B$ domain is proportional to the extremum cross-section ($S_F$) of Fermi surface, $F$=$\frac{\hbar}{2\pi e} S_F$. The Fermi surfaces of PrBi calculated by the DFT are displayed in Fig.~\ref{Fig2}(a). It consists of two hole pockets ($\beta$ and $\gamma$) co-centered at $\mathbf{\Gamma}$, and three electron pockets ($\alpha$) located at $\mathbf{X}$. Following the same notation in LaBi\cite{Hasegawa-LaBiFS}, the two fundamental frequencies $F$=255 and 560 T are assigned as (vertical) $\alpha$ and $\beta$ pocket, respectively. In addition, there is a small peak at 1511 T. By comparing with LaBi\cite{Hasegawa-LaBiFS}, this frequency should originate from those horizontal $\alpha$ pockets whose elongate axes are in $\mathbf{k_x}$ and $\mathbf{k_y}$ so that their cross-sections are much larger. We, therefore, denote it $F_{\alpha'}$. These SdH results are comparable with Ref.~\cite{VashistA-PrBiSdH}. The structure of the Fermi surfaces can be mapped out by angular dependent SdH measurements [Fig.~\ref{Fig2}(e)]. As expected, $F_{\beta}$ is essentially independent on field angle, consistent with a nearly spherical shape. $F_{\alpha}$, in contrast, changes slowly for small angle, but increases rapidly for angles larger than 45 $^{\circ}$, in agreement with the calculated ellipsoidal character. It should be pointed out that $F_{\alpha'}$ is hardly seen between 70 and 85 $^{\circ}$, similar as the case in LaBi\cite{Hasegawa-LaBiFS}. Moreover, the star-shaped $\gamma$ pocket is not visible in our SdH measurements for all the field angle. A comparison between calculated and experimental SdH frequencies can be found in Fig.~\ref{Fig2}(b-c).

The electron carrier density can be calculated by estimating the total volume of $\alpha$ pockets, and this yields $n_e$=$4.08\times10^{20}$ cm$^{-3}$. Note that we have incorporated the spin degeneracy. Since $F_{\gamma}$ is missing in SdH, we can not obtain the hole carrier density directly, however, $n_h$ is assumed to be close to $n_e$ in semimetals\cite{Singleton-Band}. Such an electron-hole compensation effect can be evidenced by fitting the Hall resistivity $\rho_{yx}(B)$ to a two-band model\cite{Singleton-Band,LuoY-WTe2Hall}. We obtain $n_h$=4.40$\times$10$^{20}$ cm$^{-3}$, $n_e$=4.21$\times$10$^{20}$ cm$^{-3}$, and the mobilities $\mu_h$=0.53 m$^2$/V$\cdot$s and $\mu_e$=0.40 m$^2$/V$\cdot$s [see Fig.~\ref{Fig1}(f)]. The total carrier density, $n_h$$+$$n_e$, is about 8.6$\times$10$^{20}$ cm$^{-3}$, or equivalently, 0.06 in each PrBi unit cell.

The topological feature can be manifested by the Landau-level (LL) diagram, as shown in the inset to Fig.~\ref{Fig1}(c). Our pulsed-field measurements up to 53 T enable us to determine the LL index much closer to the quantum limit in good accuracy. Since $\rho_{xx}$$\gg$$\rho_{yx}$, $\rho_{xx}$ is in-phase with conductivity $\sigma_{xx}$, the minima of $\Delta\rho_{xx}$ should be assigned as half-integer Landau level\cite{Xiang-BiTeClSdH}; this criterion is in line with the classic 2D quantum Hall effect\cite{Klitzing-QHE}. The extrapolation of $\alpha$-pocket LL diagram to the infinitely large field limit results in an intercept 0.05(4). For a three dimensional Fermi surface, this intercept corresponds to $1/2$$-$$\Phi_B/2\pi$$-$$\delta$\cite{Murakawa-BiTeISdH}, where $\Phi_B$ is the Berry phase, and $-$1/8$\leq$$\delta$$\leq$1/8 is an additional phase shift due to the curvature of the Fermi surface topology. The obtained intercept 0.05(4) falling between $\pm1/8$ thus manifests a non-trivial Berry phase $\Phi_B$=1/2. Recent DFT calculations on PrBi also suggested bulk band inversion and gapless surface state, further lending support to a topologically non-trivial nature\cite{WuZ-PrSmSbBiPRB2019}.

\begin{figure}[t]
\vspace*{-15pt}
\hspace*{-13pt}
\includegraphics[width=9.5cm]{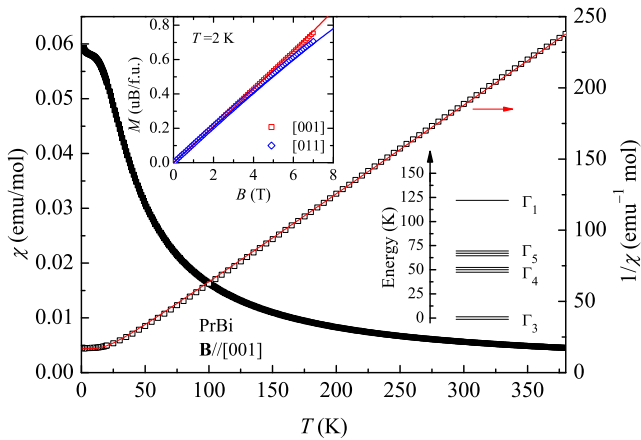}
\vspace*{-25pt}
\caption{\label{Fig3} Temperature dependence of magnetic susceptibility of PrBi measured under $B$=0.1 T and $\mathbf{B}$$\parallel$[001]. The inverse of $\chi$ is shown in the right frame. The symbols stand for experimental data, while the solid lines are theoretically calculated results based on CEF theory. The lower inset depicts the CEF splitting of the $j$=4 multiplet. The upper inset displays the field dependent isothermal magnetization at 2 K for $\mathbf{B}$$\parallel$[001] (red squares) and $\mathbf{B}$$\parallel$[011] (blue diamonds), respectively.}
\end{figure}

The main frame of Fig.~\ref{Fig3} shows the magnetic susceptibility ($\chi$) of PrBi measured with $\mathbf{B}$$\parallel$[001]. Above 25 K, $\chi(T)$ conforms to the standard Curie-Weiss law $\chi(T)$=$C/(T-\theta_W)$. The derived effective moment is 3.57 $\mu_B$, very close to the value for a free Pr$^{3+}$ ion. The fitted Weiss temperature is negligibly small, $\theta_W$=$-$0.5(2) K, much smaller than those in other cubic Pr intermetallic compounds, like PrMg$_3$ ($-$40 K)\cite{Tanida-PrMg3}, PrOs$_4$Sb$_{12}$ ($-$16 K)\cite{BauerE-PrOs4Sb12}, PrTi$_2$Al$_{20}$ ($-$40 K)\cite{Sakai-PrTrAl20JPSJ}, PrV$_2$Al$_{20}$ ($-$55 K)\cite{Sakai-PrTrAl20JPSJ}. Below 25 K, $\chi(T)$ gradually deviates and tends to saturate, suggestive of a van-Vleck paramagnetic ground state which is a consequence of CEF splitting.

The CEF Hamiltonian for Pr$^{3+}$ in $O_h$ point group (cubic) writes\cite{Bauer-CEF,Takegahara-CEFCubic}
\begin{equation}
\hat{H}_{CEF}=B_4(\hat{O}_4^0+5\hat{O}_4^4)+B_6(\hat{O}_6^0-21\hat{O}_6^4),
\end{equation}
where $\hat{O}_{l}^m$ ($l$=4,6; $m$=0,4) are Stevens operators\cite{Stevens-Operators}, and $B_4$ and $B_6$ are CEF parameters that can be determined experimentally. The nine-degenerated $j$=4 multiplet of Pr$^{3+}$ in such a CEF splits into one singlet ($\Gamma_1$), one doublet ($\Gamma_3$) and two triplets ($\Gamma_4$ and $\Gamma_5$)\cite{LLW-CEFCubic}. It is particularly interesting when the $\Gamma_3$ doublet is the ground states, in which case the correlation effect involving quadrupolar degrees of freedom are deemed to be responsible for many emergent quantum phenomena\cite{Cox-Quadrupolar}.

\begin{table*}
\caption{CEF parameters, energy levels, wave functions and the $4f$ charge distributions in PrBi at zero magnetic field. Calculated with $B_4$=0.0012(2) K, and $B_6$=$-$0.00065(3) K (or equivalently, $W$=$-$0.8911 K and $x$=$-$0.0808 in LLW parameters\cite{LLW-CEFCubic}). The quadrupolar moment operators are $\hat{O}_2^0$=$(3\hat{J}_z^2-\mathbf{J}^2)/2$, and $\hat{O}_2^2$=$\sqrt{3}(\hat{J}_x^2-\hat{J}_y^2)/2$. }
\label{Tab1}
\begin{ruledtabular}
\begin{center}
\def\temptablewidth{2.0\columnwidth}
\begin{tabular}{ccccccccccc}
$E$(K)                        &   0     &     0   &   50(5)  &   50(5) &  50(5)  &  67(5)  &   67(5) &  67(5)  &  122(5)    \\ \hline
$\mid 4,+4 \rangle$           &  0.0000 & -0.5401 &  0.0000  &  0.0000 & -0.7071 &  0.0000 &  0.0000 &  0.0000 &  0.4564    \\
$\mid 4,+3 \rangle$           &  0.0000 &  0.0000 & -0.3236  & -0.1425 &  0.0000 &  0.0000 &  0.6677 &  0.6551 &  0.0000    \\
$\mid 4,+2 \rangle$           & -0.7071 &  0.0000 &  0.0000  &  0.0000 &  0.0000 &  0.0000 &  0.4952 & -0.5047 &  0.0000    \\
$\mid 4,+1 \rangle$           &  0.0000 &  0.0000 & -0.3771  &  0.8561 &  0.0000 & -0.3536 &  0.0000 &  0.0000 &  0.0000    \\
$\mid 4,~~0 \rangle$          &  0.0000 &  0.6455 &  0.0000  &  0.0000 &  0.0000 &  0.0000 &  0.0000 &  0.0000 &  0.7638    \\
$\mid 4,-1 \rangle$           &  0.0000 &  0.0000 & -0.8561  & -0.3771 &  0.0000 &  0.0000 & -0.2524 & -0.2476 &  0.0000    \\
$\mid 4,-2 \rangle$           & -0.7071 &  0.0000 &  0.0000  &  0.0000 &  0.0000 &  0.0000 & -0.4952 &  0.5047 &  0.0000    \\
$\mid 4,-3 \rangle$           &  0.0000 &  0.0000 & -0.1425  &  0.3236 &  0.0000 &  0.9354 &  0.0000 &  0.0000 &  0.0000    \\
$\mid 4,-4 \rangle$           &  0.0000 & -0.5401 &  0.0000  &  0.0000 &  0.7071 &  0.0000 &  0.0000 &  0.0000 &  0.4564    \\ \hline
$\langle \hat{J}_z \rangle$   &  0.0000 &  0.0000 &  -0.3375 &  0.3375 &  0.0000 & -2.5000 &  1.2738 &  1.2262 &  0.0000    \\
$\langle \hat{O}_2^0 \rangle$ & -4.0000 &  4.0000 &  -7.0000 & -7.0000 & 14.0000 &  2.0000 & -0.9428 & -1.0572 & -0.0000    \\
$\langle \hat{O}_2^2 \rangle$ &  0.0000 &  0.0000 &  -8.9452 & -8.9452 &  0.0000 &  0.0000 &  0.0000 & -0.0000 &  0.0000    \\ \hline
$\hat{\rho}_{4f}(\mathbf{r})$ &
\begin{minipage}{0.1\textwidth}    \includegraphics[width=1.3cm]{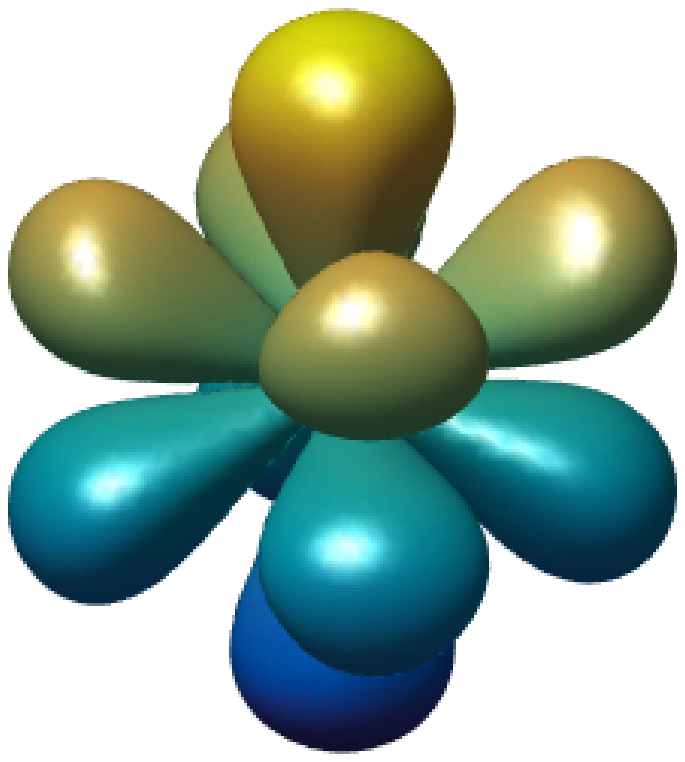}    \end{minipage}  &
\begin{minipage}{0.1\textwidth}    \includegraphics[width=1.3cm]{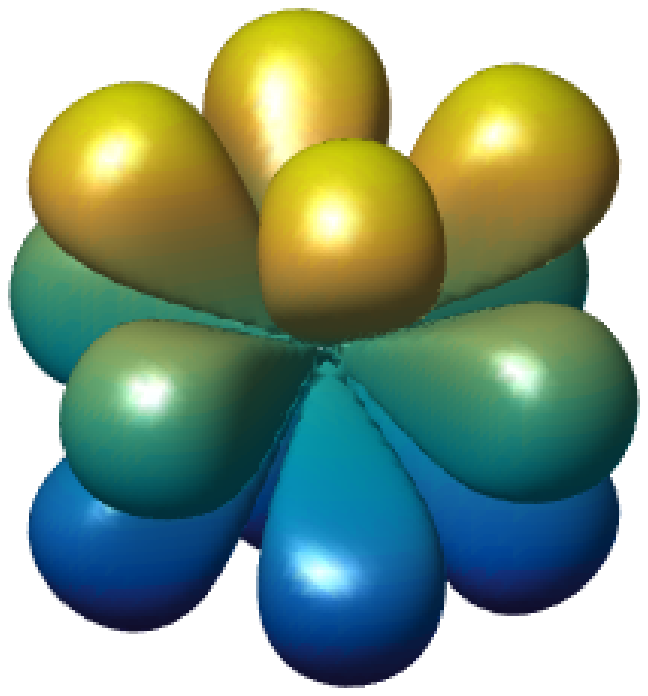}    \end{minipage}  &
\begin{minipage}{0.1\textwidth}    \includegraphics[width=1.3cm]{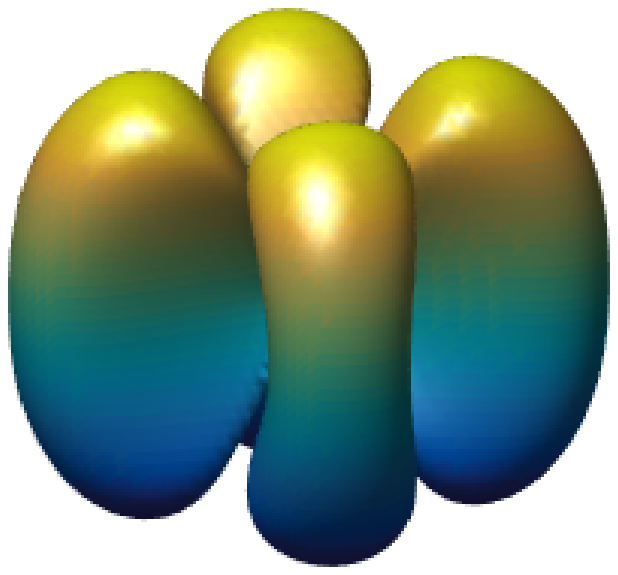}    \end{minipage}  &
\begin{minipage}{0.1\textwidth}    \includegraphics[width=1.3cm]{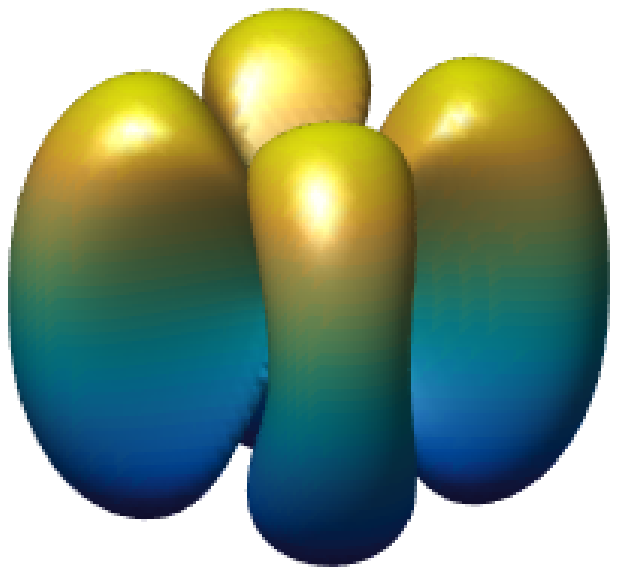}    \end{minipage}  &
\begin{minipage}{0.1\textwidth}    \includegraphics[width=1.3cm]{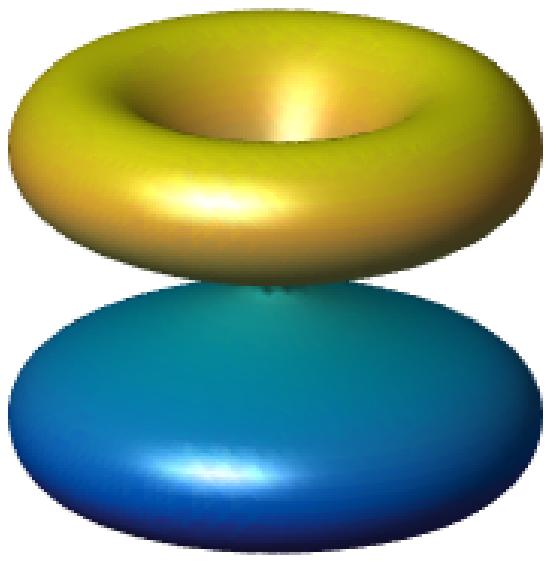}    \end{minipage}  &
\begin{minipage}{0.1\textwidth}    \includegraphics[width=1.3cm]{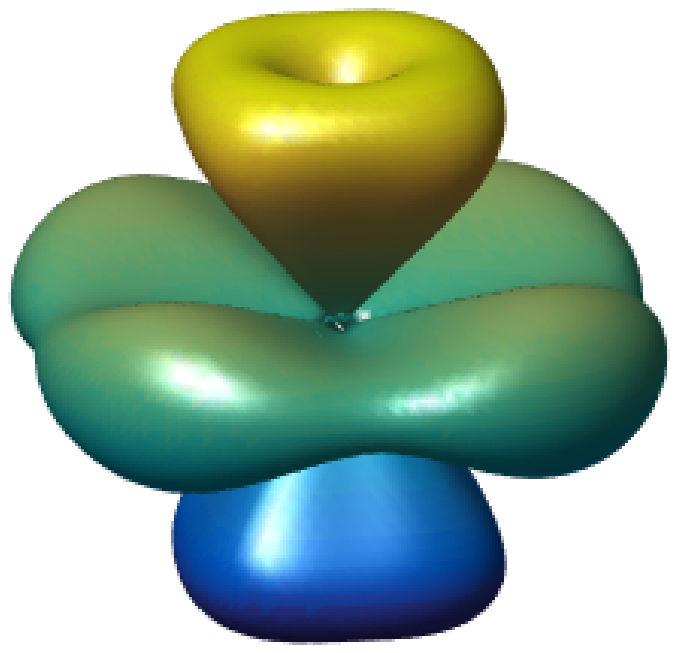}    \end{minipage}  &
\begin{minipage}{0.1\textwidth}    \includegraphics[width=1.3cm]{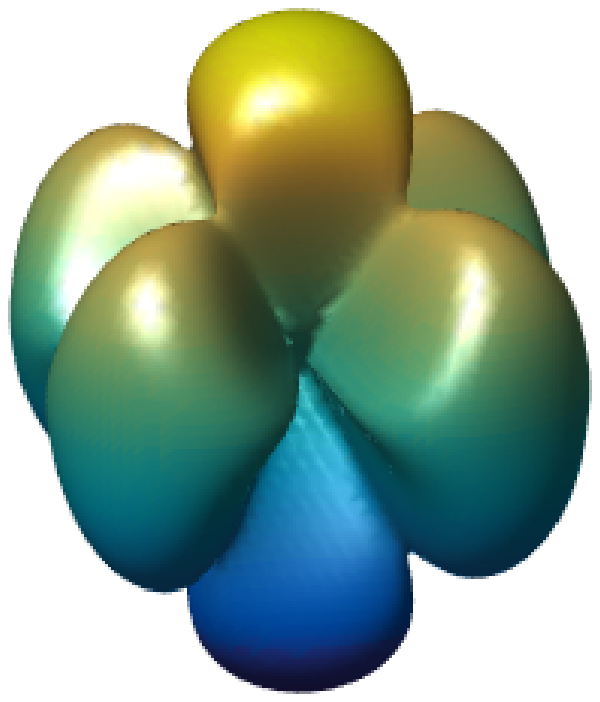}    \end{minipage}  &
\begin{minipage}{0.1\textwidth}    \includegraphics[width=1.3cm]{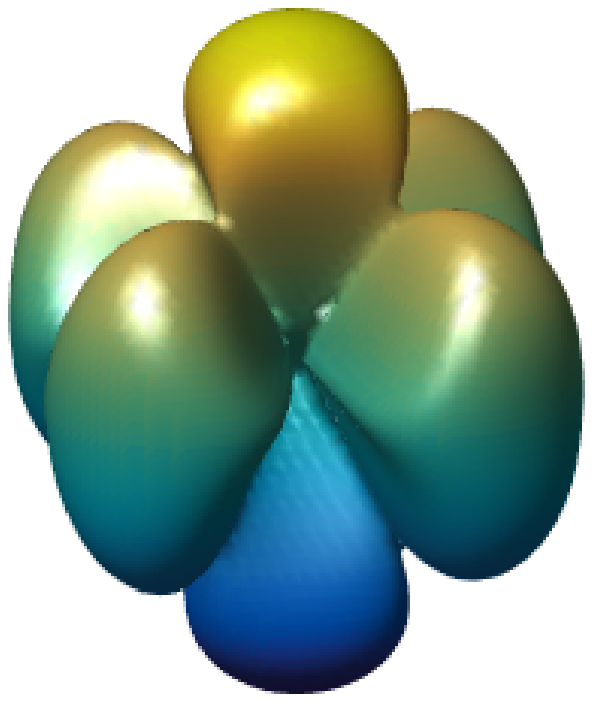}    \end{minipage}  &
\begin{minipage}{0.1\textwidth}    \includegraphics[width=1.3cm]{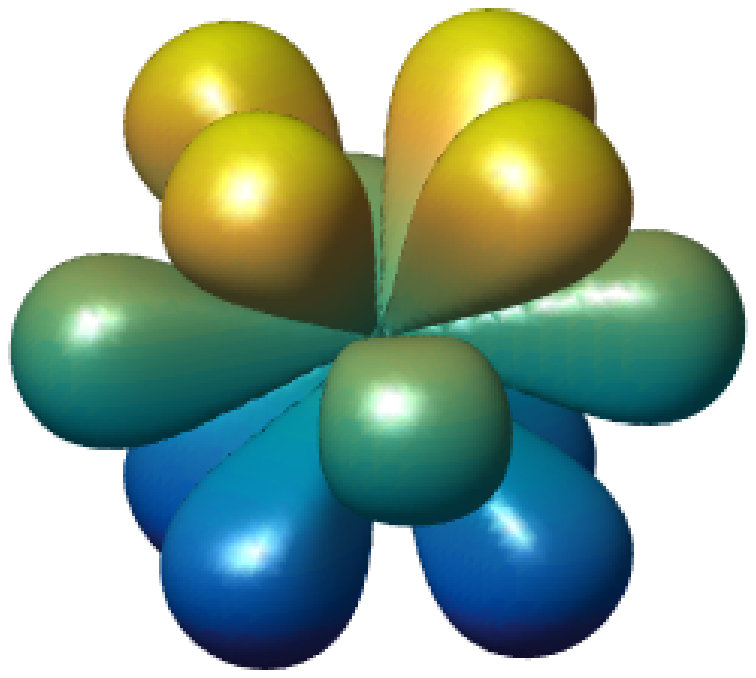}    \end{minipage}  \\
\end{tabular}
\end{center}
\end{ruledtabular}
\end{table*}

We demonstrate that $\Gamma_3$ is the ground state in PrBi by fitting the magnetic susceptibility to the CEF theory, in which $B_4$ and $B_6$ are set as free parameters. The best fitting parameters are $B_4$=0.0012(2) K, and $B_6$=$-$0.00065(3) K. This is equivalent to $W$=$-$0.8911 K and $x$=$-$0.0808 in Lea-Leask-Wolf (LLW) parameters\cite{LLW-CEFCubic}. This fitting confirms $\Gamma_3$ as the ground doublet, and the first excited state is $\Gamma_4$ triplet that sits at $\sim$50 K above $\Gamma_3$, cf the lower inset to Fig.~\ref{Fig3}. The energy level, wave function and the $4f$ charge distribution of each state are summarized in Table \ref{Tab1}. The two ground states can be expressed in terms of $\Gamma_3^{(1)}$=$-\sqrt{\frac{1}{2}}|+2\rangle -\sqrt{\frac{1}{2}}|-2\rangle$, and $\Gamma_3^{(2)}$=$-\sqrt{\frac{7}{24}}|+4\rangle +\sqrt{\frac{5}{12}}|0\rangle -\sqrt{\frac{7}{24}}|-4\rangle$. We further calculated the dipolar moment ($\langle \hat{J}_z \rangle$) and quadrupolar moments ($\langle \hat{O}_2^0 \rangle$ and $\langle \hat{O}_2^2 \rangle$) carried by each CEF state, where $\hat{O}_2^0$=$(3\hat{J}_z^2-\mathbf{J}^2)/2$, and $\hat{O}_2^2$=$\sqrt{3}(\hat{J}_x^2-\hat{J}_y^2)/2$. Both $\Gamma_3^{(1)}$ and $\Gamma_3^{(2)}$ have zero $\langle\hat{J}_z\rangle$ and $\langle\hat{O}_2^2\rangle$, but non-zero $\langle\hat{O}_2^0\rangle$. This manifests that the ground doublet is nonmagnetic (i.e., non-Kramers) in nature, and is consistent with the fact that $\chi(T)$ levels off at low temperature. The $\Gamma_3$ doublet ground state with non-zero $\langle\hat{O}_2^0\rangle$ can be further evidenced by the anisotropic field dependent isothermal magnetization $M(B)$ as shown in the upper inset to Fig.~\ref{Fig3}. Note that the magnetic response to an external field generally is isotropic for a cubic system. The observed anisotropy is because the $\langle\hat{O}_2^0\rangle$ order parameter gives rise to an additional field-induced $\langle\hat{J}_z\rangle$ component under $\mathbf{B}$$\parallel$[001]; such an induced component is expected to be weaker for $\mathbf{B}$$\parallel$[011], and should be absent for the $\langle\hat{O}_2^2\rangle$ order parameter with any field direction\cite{Sato-PrTi2Al20INS}. Similar anisotropy was also seen in other Pr-based intermetallic compounds such as PrIr$_2$Zn$_{20}$ \textit{et al}\cite{Onimaru-PrIr2Zn20AFQSC,Onimaru-PrRh2Zn20AFQSC}. This CEF analysis also derives a small exchange field parameter $\lambda$=$-$0.11 mol/emu, which agrees well with the small $\theta_W$ and is suggestive of weak exchange interaction between local moments.

\begin{figure}[t]
\vspace*{-15pt}
\hspace*{-13pt}
\includegraphics[width=9.5cm]{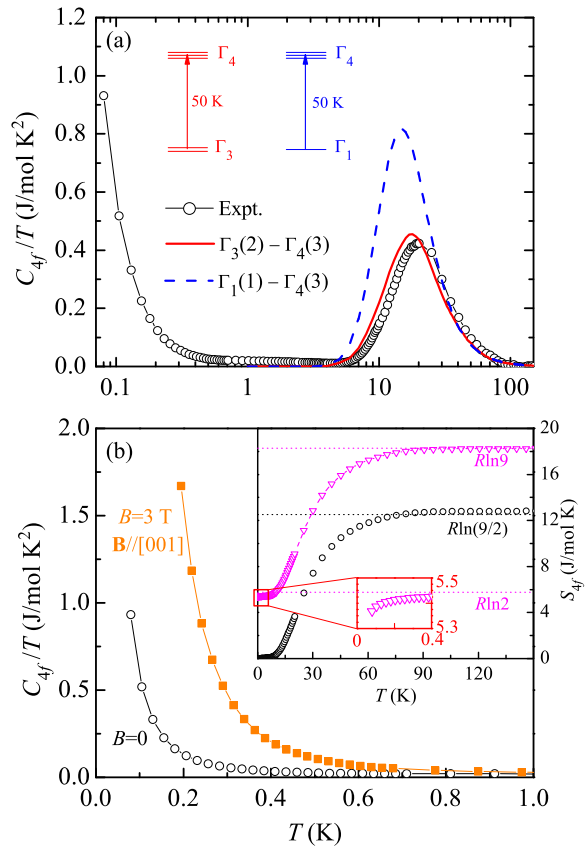}
\vspace*{-25pt}
\caption{\label{Fig4} (a) Temperature dependence of $C_{4f}/T$, the $4f$ contribution to specific heat. The red-solid and blue-dash lines are calculated Schottky anomalies based on two distinctive CEF splittings (see the inset). (b) The $C_{4f}/T$ of PrBi at low temperature, under $B$=0 and 3 T. The inset shows the integrated entropy $S_{4f}$; the black circles are calculated by assuming $S_{4f}$(2K)=0, and this leads to $S_{4f}$$\approx$$R\ln(9/2)$ at high $T$; the magenta triangles represent an inferred curve with an assumption that the degeneracy of $\Gamma_3$ doublet is further lifted by a low-$T$ quadrupolar ordering, so that $S_{4f}$ exhibits a plateau of $R\ln2$ and finally reaches $R\ln9$ at high $T$. }
\end{figure}

In Fig.~\ref{Fig4}, we display the specific heat of PrBi. To start with, the specific heat of the non-$4f$ analog LaBi (shown in Fig.~\ref{FigA1}) at low temperature obeys the classic formula $C_{La}$=$\gamma_{La} T$$+$$\beta_{La} T^3$, with Sommerfeld coefficient $\gamma_{La}$$\approx$1 mJ/mol$\cdot$K$^2$. Such a small $\gamma_{La}$ is in consistency with the semimetallic nature. After subtracting $C_{La}$ from the total specific heat of PrBi, we obtain the contribution from $4f$ electrons, $C_{4f}$. We show $C_{4f}/T$ as a function of $T$ in Fig.~\ref{Fig4}(a). A broad peak can be seen at around 25 K, which should be attributed to the Schottky anomaly due to CEF splitting. The peak position also suggests that the first excited level should be about 50 K above the ground state. We simulated the Schottky anomaly specific heat by considering $\Gamma_3$ or $\Gamma_1$ (both are nonmagnetic) as ground state, and the results are depicted as the red-solid and blue-dash lines in Fig.~\ref{Fig4}(a), respectively. This confirms the $\Gamma_3$ doublet as the ground state. We then derived the entropy by integrating $C_{4f}/T$ over $T$ from 2 K to 150 K. The entropy gain saturates at $R\ln(9/2)$ at high temperature, and this provides further evidence for $\Gamma_3$ doublet as the CEF ground state, and moreover, this degeneracy remains un-lifted down to 2 K. Additional discussion about the Pr$^{3+}$ ground state can be found in \textbf{Appendix}. We did not see the $R\ln(5/2)$ plateau, because the energy levels of $\Gamma_4$ and $\Gamma_5$ are too close.

It is possible that the degeneracy of the $\Gamma_3$ doublet ground state can be further lifted by quadrupolar Kondo effect or forming a quadrupolar ordering at lower temperature. To clarify this issue, we measured the specific heat down to 0.08 K with Dilution Refrigerator. A striking feature is that $C_{4f}/T$ increases rapidly below 0.1 K. Such an increase does not conform to a $-\log T$ or $T^{-3}$ behavior, which excludes Kondo effect or nuclear Schottky anomaly as an origin\cite{Cox-Quadrupolar,Gopal-SpeciHeat}. Actually, both $^{141}$Pr (nuclear spin $^{141}I$=5/2) and $^{209}$Bi ($^{209}I$=9/2) sit in local cubic-site symmetry with vanishing electric field gradient, the zero-field nuclear Schottky anomalies of them should be pushed to $T$=0. Since the CEF ground state is a nonmagnetic $\Gamma_3$ doublet, magnetic transitions are not likely, but rather quadrupolar transitions. Under magnetic field $\mathbf{B}$$\parallel$[001], this anomaly moves to higher temperature [cf Fig.~\ref{Fig4}(b)]. This rules out a superconducting transition but is compatible with the quadrupolar ordering as seen in Pr$Tm_2X_{20}$\cite{Sakai-PrTrAl20JPSJ,Onimaru-PrIr2Zn20AFQSC}. Note that an external magnetic field along [001] induces a dipolar component that can promote a quadrupolar ordering\cite{Shiina-CeB6Field}. However, we should also mention that this field effect on the low-$T$ specific heat probably have been exaggerated by the hyperfine-enhanced nuclear Schottky anomaly\cite{Gopal-SpeciHeat}. Because this quadrupolar phase transition is just below the base temperature of our measurements, we are not able to calculate the entropy. An inferred $S_{4f}(T)$ at zero field can be given that it restores $\sim$$R\ln2$ above the transition temperature and finally will saturate to $R\ln9$ at high temperature, see the inset to Fig.~\ref{Fig4}(b). To achieve the $R\ln2$ entropy gain, the jump in $C_{4f}/T$ has to be extremely large at the transition (Note \cite{Note1}). Clearly, the low-temperature specific heat of PrBi requires further investigations.

To summarize the important findings:\\
(1) By ultrahigh field transport and quantum-oscillation measurements, we confirm PrBi as a topological semimetal with low carrier density, $n_e$$+$$n_h$$\approx$8.6$\times$10$^{20}$/cm$^{3}$, viz 0.06 /f.u.;\\
(2) The Pr$^{3+}$ $4f^2$ sitting in the cubic-symmetry CEF has a nonmagnetic $\Gamma_3$ doublet ground state that carries non-zero $\langle\hat{O}_2^0\rangle$ quadrupolar moments. They are expected to be ordered below 0.08 K;\\
(3) Both quadrupolar Kondo effect and RKKY interaction seem to be rather weak in PrBi.

Expanding further on point (3) above. The weak RKKY interaction between the quadrupolar moments can be seen not only from the super low quadrupolar transition temperature, but also indicated by the small Weiss temperature $\theta_W$ and exchange-field parameter $\lambda$. As RKKY interaction relies on conduction electrons, it is reasonable that it is weakened in the case of low carrier density. Kondo effect is reduced, as well. 0.06 /f.u. means 6 conduction electrons are to separately screen 100 quadrupolar moments, which apparently is not enough\cite{Nozieres-EPJB1998}. In PrBi, although quadrupolar Kondo effect and RKKY interaction are both reduced, the latter appears to slightly win, which leads to the quadrupolar ordering as the ground state.

Topology, thus, meets active quadrupolar degrees of freedom in this low-carrier-density semimetal. It opens a new window to the physics of topology and strongly correlated effect with quadrupolar degrees of freedom. One interesting question is how does the physical property evolve with the $c$-$f$ hybridization. We should point out that the Pr$Pn$ family ($Pn$=P,As,Sb,Bi) all crystallize in cubic structure and are all likely of $\Gamma_3$ CEF ground state. From Bi to P, the $c$-$f$ hybridization should increase due to the ``chemical pressure" effect, the low-temperature property may change. But meanwhile, the spin-orbit coupling reduces, and the topological feature probably fades out. This trend has been seen in the non-$4f$ analog La$Pn$\cite{Yang-LaAsTrivial,NiuXH-LaBiARPES}. Another means is physical pressure effect, the advantages of which are to maintain the spin-orbit coupling and to introduce little impurity or disorder. According to the well-known Doniach phase diagram\cite{Doniach}, the Kondo effect increases faster than the RKKY interaction as $c$-$f$ hybridization strengthens, we expect the quadrupolar ordering to be suppressed, quadrupolar fluctuations to proliferate as approaching some critical point, and more intriguing phenomena like heavy-fermion effect, non-Fermi liquid and/or unconventional superconductivity may emerge. Moreover, it is worthwhile to emphasize that low carrier density itself affects the character of quantum critical points. Our previous study on the (dipolar) Kondo semimetal CeNi$_{2-\delta}$As$_2$ tuned by pressure effect has manifested that unconventional Kondo-breakdown type quantum critical point\cite{Coleman-HFDimension,SiQ-localQCP} is more likely to take place in the low carrier density limit\cite{LuoY-CeNi2As2Pre}. Apart from these, it is also of great interest to see whether the topological feature switches if the CEF ground state is changed, e.g. tuned by uniaxial stress. We look forward to these peculiar phenomena in Pr-based semimetals. More experimental and theoretical works are needed in the future.

\section{\Rmnum{4}. Conclusions}

In all, the topological semimetal PrBi provides an interesting paradigm of quadrupolar degrees of freedom in the limit of low carrier density, evoking the necessity to revisit the Nozi\`{e}res exhaustion problem in the context of multi-channel Kondo effect.

\textit{Note added:} When completing this work, we became aware of several publications on the MR and SdH effect of PrBi\cite{VashistA-PrBiSdH,WuZ-PrSmSbBiPRB2019}.\\

\section{Acknowledgments}

Y. Luo thanks Roman Movshovich, Joe D. Thompson and Stuart Brown for helpful discussions. This work is supported by the National Science Foundation of China (Grant nos. 51861135104, 11574097 and 11874137),  National Key Research and Development Program of China (Grant no. 2016YFA0401704), Fundamental Research Funds for the Central Universities (Grant no. 2019kfyXMBZ071) and China Postdoctoral Science Foundation (2018M630846). Y. Li is supported by National Natural Science Foundation of China (Grant No. U1932155). S. Wang, Z. Zhu and Y. Luo acknowledge 1000 Youth Talents Plan of China.

\setcounter{table}{0}
\setcounter{figure}{0}
\setcounter{equation}{0}
\renewcommand{\thefigure}{A\arabic{figure}}
\renewcommand{\thetable}{A\arabic{table}}
\renewcommand{\theequation}{A\arabic{equation}}

\section{Appendix: A\lowercase{dditional results and discussion about specific heat}}

\begin{figure}[t]
\vspace*{-15pt}
\hspace*{-13pt}
\includegraphics[width=9.5cm]{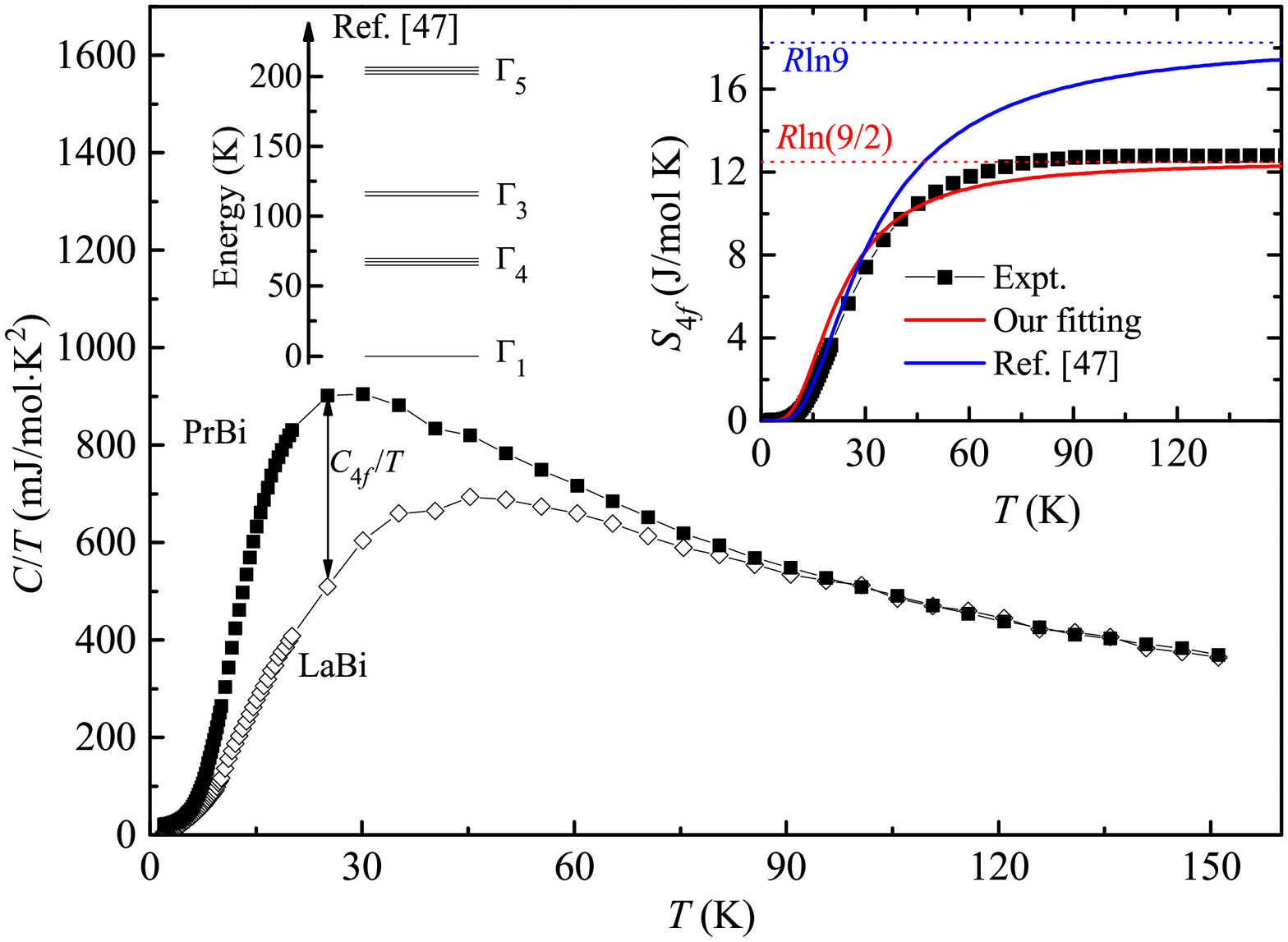}
\vspace*{-25pt}
\caption{\label{FigA1} Main frame, temperature dependence of $C/T$ of LaBi (circles) and PrBi (squares). The inset displays magnetic entropy gain based on different CEF splittings: $\Gamma_3$ doublet ground state (red) and $\Gamma_1$ singlet ground state (blue), the latter of which is simulated with the parameters given by Birgeneau {\it et al} \cite{Birgeneau-PrBiNeutron}. Note that here the integration for the experimental results is taken between 2 K and 150 K.}
\end{figure}

Figure~\ref{FigA1} displays the raw data of $C/T$ of LaBi (open diamonds) and PrBi (solid squares), as functions of temperature. The difference between them yields $C_{4f}/T$, as shown in Fig.~\ref{Fig4}.

We notice that some earlier papers in 1970s claimed $\Gamma_1$ as ground state\cite{Birgeneau-PrBiNeutron} and a possible nuclear ordering below 10 mK in PrBi\cite{Andres-PrBiC}. While the latter is far below the base temperature of our measurements, the former is not consistent with the magnetic entropy. This is because if $\Gamma_1$ is the ground state, the entropy gain should approach $R\ln 9$ at high temperature, as depicted by the blue line in the inset to Fig.~\ref{FigA1}. This is apparently different from our experimental data.

%

\end{document}